# Super-resolution imaging through a multimode fiber: the physical upsampling of speckle-driven

Chuncheng Zhang[1], Tingting Liu[1], Zhihua Xie[2], Yu Wang[1], Tong Liu[1], Qian Chen[1], Xiubao Sui[1*]

1. School of Electronic Engineering and Optoelectronic Technology, Nanjing University of Science and Technology, Nanjing 210094, China
2. Key Lab of Optic-Electronic and Communication, Jiangxi Science and Technology Normal University, Nanchang 330013, China

*Corresponding email: sxb@njust.edu.cn

Following recent advancements in multimode fiber (MMF), miniaturization of imaging endoscopes has proven crucial for minimally invasive surgery in vivo. Recent progress enabled by super-resolution imaging methods with a data-driven deep learning (DL) framework has balanced the relationship between the core size and resolution. However, most of the DL approaches lack attention to the physical properties of the speckle, which is crucial for reconciling the relationship between the magnification of super-resolution imaging and the quality of reconstruction quality. In the paper, we find that the interferometric process of speckle formation is an essential basis for creating DL models with super-resolution imaging. It physically realizes the upsampling of low-resolution (LR) images and enhances the perceptual capabilities of the models. The finding experimentally validates the role played by the physical upsampling of speckle-driven, effectively complementing the lack of information in data-driven. Experimentally, we break the restriction of the poor reconstruction quality at great magnification by inputting the same size of the speckle with the size of the high-resolution (HR) image to the model. The guidance of our research for endoscopic imaging may accelerate the further development of minimally invasive surgery.

## 1. Introduction

Optical fibers serve as a crucial component in numerous cutting-edge applications within the biomedical field, including laser therapy[1, 2], spectroscopic analysis[3, 4], and endoscopic imaging[5, 6]. By leveraging the unique transmission properties in optical fiber, endoscopes enable real-time visualization of patients' internal tissues during surgeries, thus enhancing surgical precision and safety[7, 8]. However, endoscopes typically use a fiber bundle composed of thousands of single-mode fibers for image transmission, which leads to large bundle sizes and low resolution[9-11]. These drawbacks not only degrade the quality of images but also negatively affect the patients' postoperative recovery. Fortunately, advances in multimode fibers(MMF) enable parallel transmission of multiple independent channels[12-15]. Hence, MMF has paved the way for the development of compact endoscopes.

Over the past several decades, various methods have been developed to achieve imaging through MMF, such as wavefront shaping[16-19], transmission matrix[20-22], and phase conjugation[23, 24]. Although these methods can address the issue of speckle patterns formed by the interference between different propagation modes in MMF, they are still limited in terms of resolution[25, 26]. Due to the small size of MMF (i.e., the small diameter of the core), the number of propagation modes is limited. Therefore, the speckle size is large, resulting in reduced spatial information obtained from detection, which subsequently leads to a low spatial resolution [27]. Based on the above, people must balance miniaturization and resolution when imaging trauma-sensitive in vivo organs such as the brain. Nowadays, deep learning(DL), which learns features of a target from the data, has become one of the essential tools for imaging through MMF with its powerful reconstruction capability [28-31]. Initially, high-fidelity images were successfully reconstructed by using a data-driven model, optimizing the model and training strategy[32-35]. Recent advancements in DL models' integration of speckles with highly redundant information, including the prior knowledge of speckle correlation theory for physics-informed learning or compressive imaging without any constraints on the sample sparsity, further enhancing the resolution[36-40]. However, the guidance of these methods for upsampling in the model is unclear, i.e., the expansion of pixels depends on the correlation between the numerical values, limiting the reconstruction details and the magnification[41, 42]. Furthermore, a large magnification leads to poor reconstruction quality[43]. Therefore, it is an important research direction to extend the upsampling capability of DL models by recovering the missing details from the physical characteristics of speckles. The limitation of the correspondence between magnification and reconstruction quality is likely to be broken.

In this paper, we investigate the physical upsampling of speckle-driven based on the interference principle of light in multimode fibers (MMF), aiming to break the curse of low image quality at large magnification in DL models. Specifically, we find that the interferometric process of speckle at the far end of the MMF achieves physical pixel expansion, compensating for the perceptual loss from the insufficient correlation in the data drive. The validity of our findings was experimentally verified. The experiment generates speckles by low-resolution (LR) images passed through the MMF. We use speckle patterns of the same size as the high-resolution (HR) images for training, which effectively addresses the lack of detail in upsampling by data drive. Moreover, thanks to the speckle redundancy, the reconstruction quality is improved after the large speckles are compressed to the same size as the HR image. In addition, our mechanism enhances the generalization ability of the DL models. Our findings will further advance the application of endoscopic imaging in minimally invasive surgery.

## 2. METHODS

## 2.1 Physical Basic

The proposed physical upsampling of speckle-driven comes from speckle interference. The speckle pattern appears chaotic, but a linear optical relation exists between MMF's near and far optical fields. Thus, the MMF can be considered an optical transformation system with a transmission matrix, where the speckle pattern is the direct mapping of the LR image through the MMF. The transformation relation between the LR image at the proximal end and the speckle pattern at the distal light field can be expressed as:

$$E(\mathbf{r}_b) = \int G(\mathbf{r}_b, \mathbf{r}_a) E(\mathbf{r}_a) d\mathbf{r}_a \tag{1}$$

Where the amplitude of the light source at the position $\mathbf{r}_a$ on the LR image is $E(\mathbf{r}_a)$, and the amplitude of the light source at the position $\mathbf{r}_b$ on the speckle is $E(\mathbf{r}_b)$. $G(\mathbf{r}_b, \mathbf{r}_a)$ is a Man-Green function describing the change of the light field from $\mathbf{r}_a$ to $\mathbf{r}_b$. However, the proximal and distal optical fields cannot be approximated as consisting of an infinite number of point sources because of the limitation of the number of pixels in the detector and the propagation mode of the MMF. In space, the definitions of $\mathbf{r}_a$ and $\mathbf{r}_b$ are not continuous. Thus, assuming that the pixels of the LR image $E_n^{in}$ is N, $G(\mathbf{r}_b, \mathbf{r}_a)$ is the optical transmission matrix $t_{mn}$ of the MMF, consisting of M × N complex-valued coefficients. The speckle pattern $E_m^{out}$ received by the detector can be expressed as:

$$E_m^{out} = \sum_n^N t_{mn} E_n^{in} \tag{2}$$

From equation (2), each pixel is highly redundant and contains all the information of the LR image, implying that the speckle pattern itself may contain richer information than the LR image. Thus, the formation of speckle patterns can be considered an evolutionary strategy, and the mapping relation physically increases the amount of information that the DL model can perceive. At the same time, the number of speckle pattern pixels is M. Assuming that M is bigger than N, the interference process of the speckle invisibly completes the upsampling of the LR image, as shown in Fig.1. The physical upsampling avoids data-driven information loss and reduces the impact on the image itself.

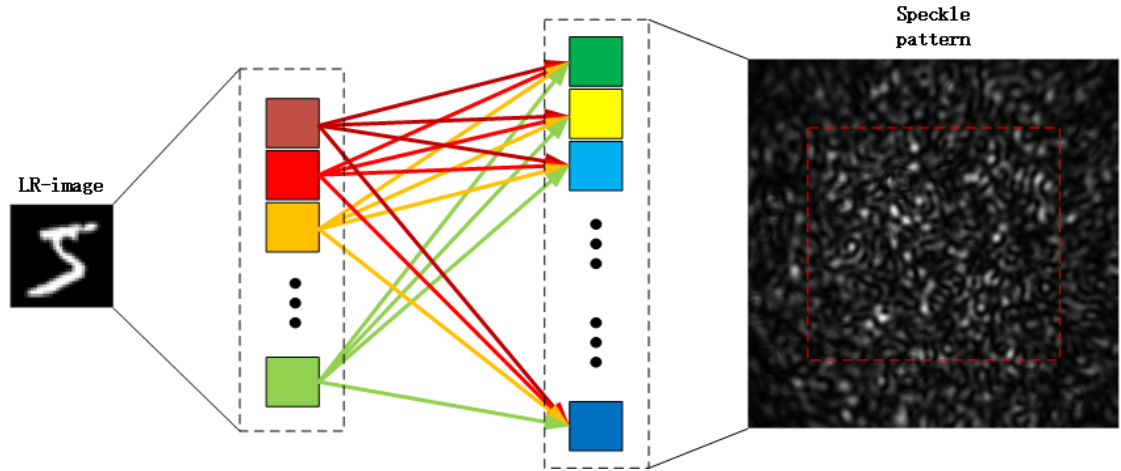

Fig.1. Formation of speckles. The size of the speckles within the red dashed line is consistent with the HR image.

## 2.2 Experimental Implementation

The experimental setup is illustrated in Fig.2. A beam from a 532 nm continuous laser (LCX-532S, Oxxius, 80 mW) is first expanded and collimated onto a spatial light modulator (SLM, PLUTO-2-vis-096, HOLOEYE, pixel pitch: 8 μm) by the beam expander (GBE20-A-20×, Thorlabs, 20x). Polarizers P is used to select the appropriate polarization state for the SLM. SLM is used to code and display the 8-bit objects. The modulated light is projected onto the objective (RMS10X, Thorlabs, NA:0.25) to be coupled into an MMF by a 4F lens system (L1:

AC254-200, Thorlabs, f: 200mm; L2: AC254-150, Thorlabs, f: 150mm). The scattered light from the MMF is collected by the other objective (RMS20X, Thorlabs, NA:0.4) and is finally detected by a CMOS (BFS-U3–04S2M-CS, Point Gray; pixel pitch: 6.9 μm).

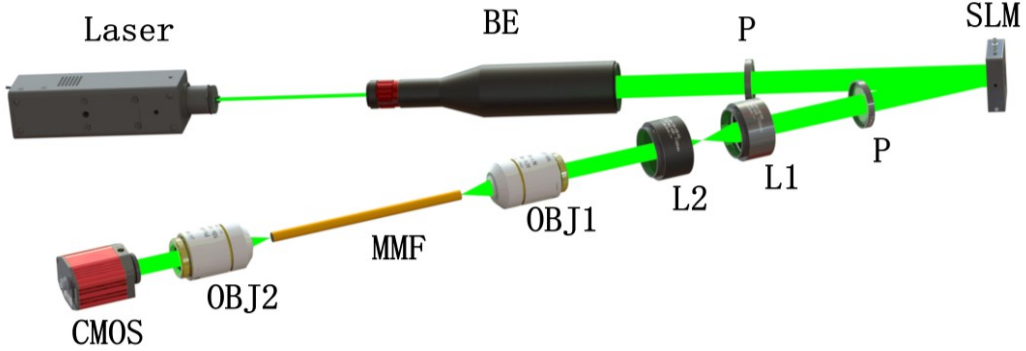

Fig.2. Schematic of the experimental setup of projecting LR-images through MMF; BE: beam expander; L1, L2: lens; P: polarizer; OBJ1 and OBJ2: objective lens; SLM: spatial light modulator; MMF: multimode fiber.

As shown in Fig.3, a well-known U-Net model, including convolutional layers, dense blocks, skip connections, Pixel Shuffle, etc. [34], is used to reconstruct HR images. Each dense block contains multiple layers that are used to improve the training efficiency, in which each layer consists of batch normalization (BN), the rectified linear unit (ReLU) nonlinear activation, and convolution. Skip connections are utilized between layers to integrate high-level semantic and low-level local detail features. In addition, with Popoff's theory of normal distribution of random scattering medium [22, 44], we replace the original up-sampling with PixelShuffle[45]. PixelShuffle, which takes full advantage of the pixel arrangement in the speckle pattern, improves the reconstruction quality[46]. Then, super-resolution imaging also requires upscaling the size and details of the image. Therefore, we use the structural similarity (SSIM) as the model's loss function to constrain the training process[47]. SSIM is a computational image quality evaluation method with high sensitivity to the perceived image quality by simultaneously considering the image's brightness, contrast, and structural information [42]. Its value is distributed from -1 to 1, meaning it is entirely uncorrelated to wholly correlated. In DL, the value of the loss is reduced to reconstruct a positive HR image. Therefore, we use $1 - \text{SSIM}(x, y)$ as the loss function:

$$\text{Loss} = 1 - \text{SSIM}(x, y) \qquad (3)$$

$$\text{SSIM}(x, y) = 1 - \frac{(2\mu_x\mu_y + c_1)(2\sigma_{xy} + c_2)}{(\mu_x^2 + \mu_y^2 + c_1)(\sigma_x^2 + \sigma_y^2 + c_2)} \qquad (4)$$

where $\mu_x$ and $\sigma_x^2$ are the averages and variances of x, respectively. $\mu_y$ and $\sigma_y^2$ are the averages and variances of y, respectively. $\sigma_{xy}$ is the covariance of x and y. $c_1 = (k_1 L)^2$ and $c_2 = (k_2 L)^2$ are the constants, maintaining stability and avoiding division by zero[48]. Usually, the value of L is 255 $(2^{Bit} - 1)$ for 8-bit objects. The general settings of $k_1$ and $k_2$ are 0.01 and 0.03, respectively, which are appropriate[49]. We use the adam algorithm to optimize the model parameters. The learning rate is set to $10^{-4}$ to ensure that the structure of encoding-decoding has a sufficient information gradient. U-net is trained with 30 epochs. Keras/Tensorflow was used for the experiments, and the GPU was GeForce RTX 3090 Ti.

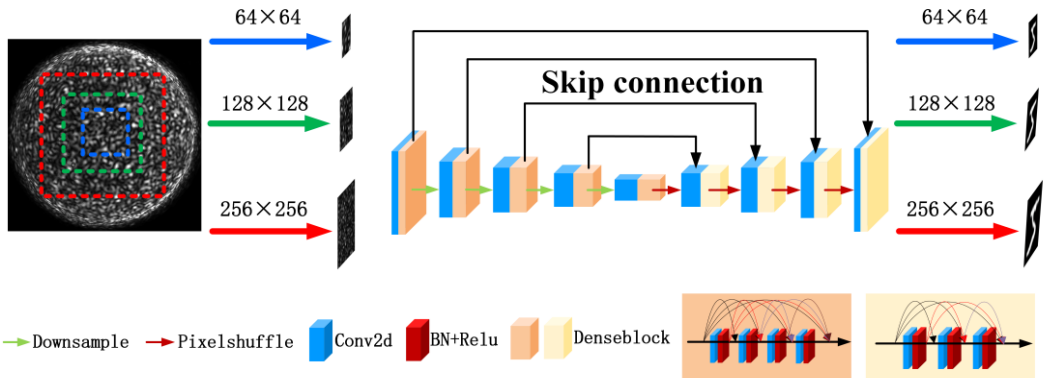

Fig.3. The physical perception of speckles for super-resolution imaging.

Our experiment uses handwritten digits from the MNIST dataset as experimental subjects. We use 4000 images to train, i.e., seen objects, and 1000 images to test, i.e., unseen objects. The bicubic interpolation down-sampling(Bi) is used to achieve degradation from HR images (256×256) to LR images (32×32). The LR image is projected as a speckle pattern by the experimental equipment in Fig 2. Next, we intercept the region of the speckle pattern with the same sizes as the HR image to achieve physical upsampling. Specifically, we intercepted 64×64, 128×128 and 256×256 speckles into U-net and reconstructed the corresponding 64×64, 128×128 and 256×256 HR images to achieve super-resolution imaging with 2x, 4x, and 8x.

## 3. RESULTS

To verify the advantages of the physical upsampling of speckle-driven, the training and testing results of the data-driven, combined data-driven, and compressed perception approaches are compared with the approach proposed in this paper. The experimental results demonstrate the excellent performance of speckle-driven physical upsampling for super-resolution and generalized imaging. In the experiment, we randomly select the reconstruction results for presentation. Pearson correlation coefficient (PCC), peak signal-to-noise ratio (PSNR), and SSIM were used to evaluate the results quantitatively.

The experimental results based on the physical upsampling of speckle-driven for seen and unseen objects are shown in Fig.4. From a subjective point of view, the reconstruction results for both seen and unseen objects are visually excellent. However, different magnifications lead to different effects of the reconstructions. The rule is as we want to show: large magnification, superior reconstruction, as shown in Figure 4(b-c). After the LR image is magnified by 8x, the topological structure and local details of both seen and unseen objects are visually close to perfect, as shown in Figure 4(a). The average PCC, SSIM, and PSNR of the seen objects are as high as 99.4%, 99.1%, and 28.3dB, respectively. The averaged PCC, SSIM, and PSNR of unseen objects are as high as 97.5%, 94.9%, and 21.1 dB, respectively. This phenomenon suggests that the physical upsampling of speckle-driven can help DL models reconstruct superior-quality HR images. At the same time, U-net shows good generalization ability. Next, at the magnification of 4, the reconstruction quality and local degradation of the seen objects appear slightly.  In unseen objects, the quality degradation of the reconstruction increases, and errors in local details become apparent. The value of their quantitative evaluation is slightly reduced than that of 8x magnification.  However, it was still acceptable. The worst reconstruction is for unseen objects at magnifications of 2.  The physical perception of speckle-driven is reduced, leading to a deterioration of the U-net's generalization ability. It is not only the local details of the digits that are not reconstructed  but also the deviations in the direction of some lines.  The values for the quantitative evaluation of all magnifications are shown in Table 1, demonstrating from the data that  speckle-driven can effectively overcome the problem of insufficient detail due to data-driven upsampling.  Physical upsampling helps super-resolution imaging overcome the limitations of magnification. Interestingly, the robustness of the U-net is further enhanced as the physical perception of the speckles is enhanced, and the standard deviation of PCC and SSIM are subsequently reduced.

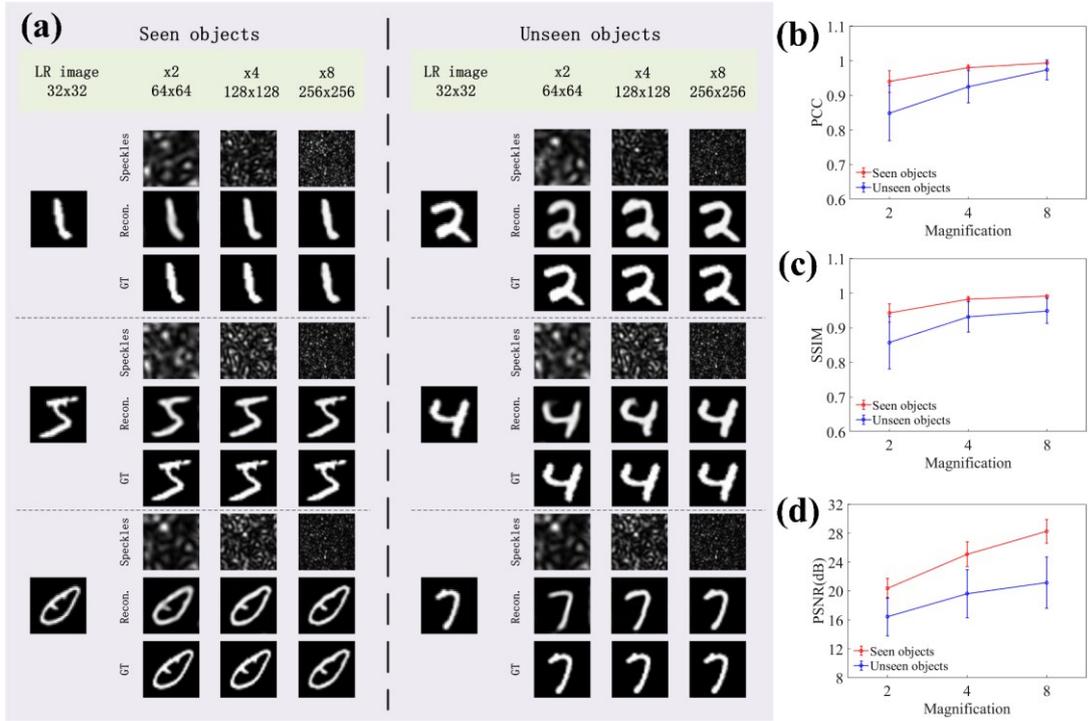

Figure 4 Results of super-resolution imaging with the physical upsampling of speckle-driven. (a) Results of imaging at magnifications of 2, 4, and 8. (b), (c), and (d) Curve chart of averaged PCC, SSIM, and PSNR with error bar, respectively.

Table 1. Quantitative evaluation of the super-resolution imaging with the physical upsampling of speckle-driven.

| Magnification | Object | PCC | SSIM | PSNR(dB) |
|---|---|---|---|---|
| x8 | Seen | 99.4% | 99.1% | 28.3 |
|  | Unseen | 97.4% | 94.9% | 21.2 |
| x4 | Seen | 98.0% | 98.2% | 25.1 |
|  | Unseen | 92.5% | 93.1% | 19.6 |
| x2 | Seen | 94.0% | 94.2% | 20.4 |
|  | Unseen | 84.9% | 85.7% | 16.5 |

To confirm the superiority of the proposed method in the paper, the data-driven approach was first selected for comparison in the experiments. The physical upsampling is replaced with upsampling of data-driven, i.e., additional up-sampling modules are added to the U-net framework. LR images are reconstructed to HR images by data-driven. We use the speckles corresponding to the LR images as the model's input at all magnifications. The size of the speckle is 32×32. Data-driven upsampling relies on correlations between pixel values, and the expansion of the number of pixels lacks guidance from the compositional logic of the target, leading to limitations of current methods for super-resolution imaging suffering from large magnifications, as shown in Fig.5. The reconstructed images at high magnification show errors in both topological structure and local details. The values of the quantitative evaluation are shown in Table 2, which are much lower than the method proposed in the paper. Taking the 2x magnification as an example, it has the worst reconstruction quality among the method of the physical upsampling of the speckle-driven and the best among the data-driven for super-resolution imaging. For the seen objects, the PCC, SSIM, and PSNR of the proposed method in the paper are 6.0%, 5.9%, and 2.9 dB higher than the data-driven method, respectively. For the unseen objects, the proposed method's PCC, SSIM, and PSNR are 7.8%, 7.5%, and 1.8 dB higher than that, respectively. Moreover, as shown in Fig. 5(b-d), the magnification limits the super-resolution imaging. The results inversely verify that DL perceives the constitutive logic of the target by super-resolution imaging with the physical upsampling of the speckle-driven are significantly better than data-driven ones.

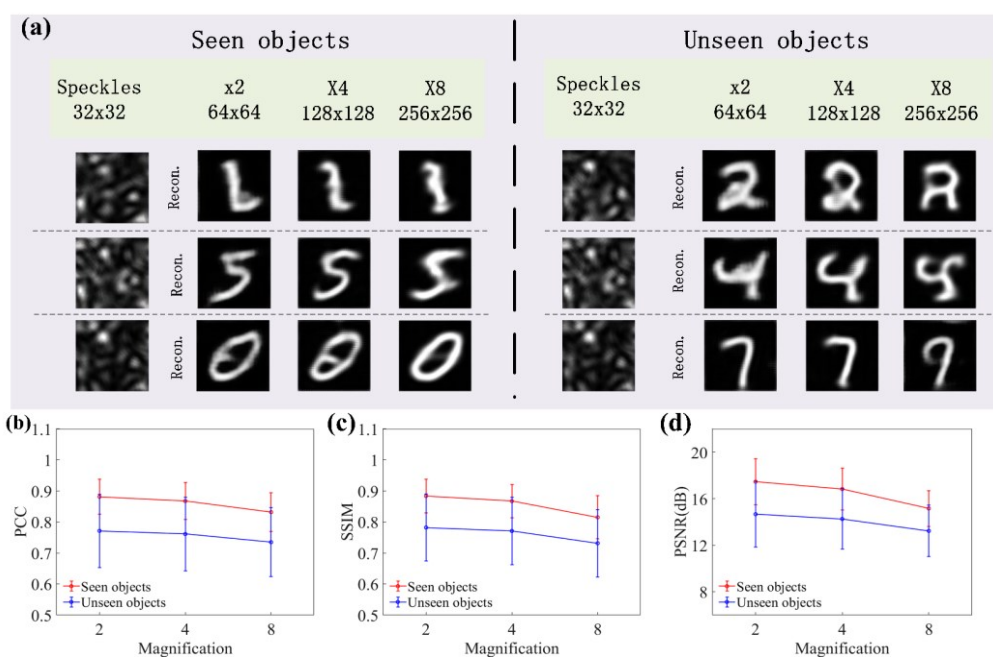

Figure 5 Results of super-resolution imaging with data-driven. (a) Results of imaging at magnifications of 2, 4, and 8. (b), (c), and (d) Curve chart of averaged PCC, SSIM, and PSNR with error bar, respectively.

Table 2. Quantitative evaluation of the super-resolution imaging with data-driven.

|  | Object | PCC | SSIM | PSNR(dB) |
|---|---|---|---|---|
| x8 | Seen | 83.2% | 81.5% | 15.2 |
|  | Unseen | 73.5% | 73.1% | 13.2 |
| x4 | Seen | 86.7% | 86.7% | 16.8 |
|  | Unseen | 76.1% | 77.1% | 14.3 |
| x2 | Seen | 88.1% | 88.4% | 17.5 |
|  | Unseen | 77.1% | 78.2% | 14.7 |

To further confirm the superiority of the method proposed in this paper, we exploit the highly redundant target information in the speckle for super-resolution imaging by coupling compressed sensing and data-driven. The 256×256 speckles are compressed into 32×32 by $B_i$ as input to U-net with additional downsampling. The quality of the reconstructed HR images is significantly improved. The degradation in reconstruction quality from the large magnification is mitigated because more information is perceived by the data-driven DL model with compressed sensing, as shown in Fig.6 and Table 3. It is worth noting that the reconstruction effect of this method is even better than the method proposed in this paper when the magnification is 2. Previous studies have confirmed the phenomenon: increasing the number of speckles can improve the amount of information perceived by the data-driven, enhancing the DL model's reconstruction capability [40]. However, at 4x and 8x magnifications, the reconstruction of compressing the speckle is not as effective as the method proposed in the paper. The lack of information in data perception eliminates some or all of the gain in information from the compression of the speckle. At the same time, the limitation of the reconstructed quality is poor when the magnification is large still exists, as shown in Fig.6(b-d). Therefore, compressed sensing can effectively improve the reconstruction quality and slow down the trend of poor imaging results with large magnification. However, at large magnifications, the advantage is not as good as the method proposed in the paper.

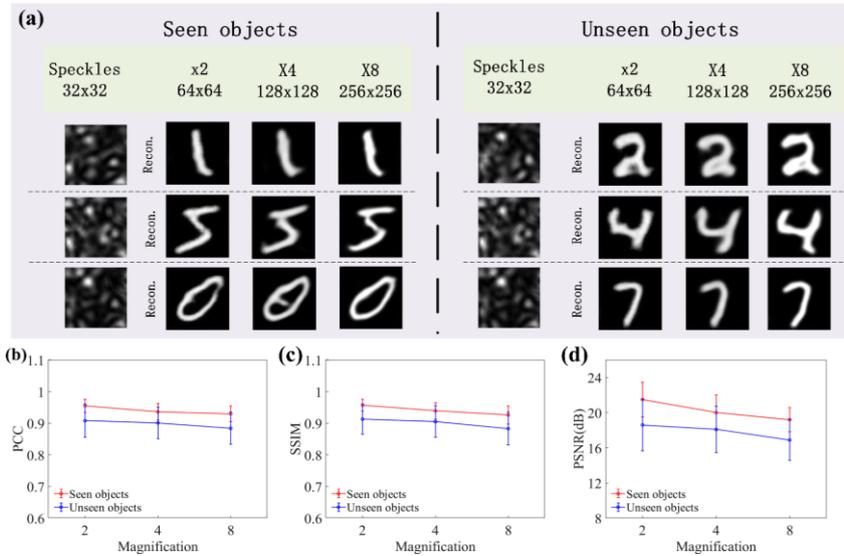

Figure 6 Results of super-resolution imaging with a combination of compressed sensing and data-driven. (a) Results of imaging at magnifications of 2, 4, and 8. (b), (c), and (d) Curve chart of averaged PCC, SSIM, and PSNR with error bar, respectively.

Table 2. Quantitative evaluation of super-resolution imaging with a combination of compressed sensing and data-driven.

|    | Object | PCC   | SSIM  | PSNR(dB) |
|----|--------|-------|-------|----------|
| x8 | Seen   | 93.0% | 92.6% | 19.2     |
|    | Unseen | 88.4% | 88.3% | 16.9     |
| x4 | Seen   | 93.6% | 93.9% | 20.0     |
|    | Unseen | 89.8% | 90.5% | 18.1     |
| x2 | Seen   | 95.4% | 95.7% | 21.5     |
|    | Unseen | 90.8% | 91.3% | 18.6     |

## 4. DISCUSSIONS

The power of the physical upsampling method proposed in the paper derives from the physical perception of speckle-driven. The following factors influence the physical upsampling: 1. high redundancy of speckle information distribution; 2. number of pixels of speckle pattern $M$. The coherent light through the different propagation modes of MMF is interfered with to form speckles with redundancy. In section 3, we demonstrate that simple compression using the speckle's redundancy can compensate for the missing information and effectively improve the reconstruction quality of super-resolution imaging. Similarly, the redundancy of the speckle is advantageous for the method of physical upsampling. Compressed sensing enhances the system's sampling rate, and the physical perception of speckle-driven is subsequently enhanced. Related experiments were used to confirm the conclusion. Firstly, the size of the complete speckle pattern captured by the camera is less than $512 \times 512$ because of the optical system. Then, to ensure the comparability of the experiments, the speckles of size 128×128 and 256×256 were compressed to 64×64 by Bi as the input of the DL model for super-resolution imaging with a magnification of 2, respectively. The reconstruction of unseen objects represents the generalization capability of the methods in the paper. In this section, we only show the reconstruction results of unseen objects, as shown in Fig.7. U-net's input contains more information than before when the high-latitude speckles are compressed to the same size as the high-resolution image. Compressed sensing is equivalent to data augmentation for the physical upsampling of speckle-driven, which ultimately improves the generalization ability of U-net, and the values of average PCC, SSIM, and PSNR are all larger than that without the compression perception, as shown in Table 4. However, there is a limit to the degree of compression of the speckles, which does not represent good reconstruction quality with a significant degree of compression. As shown in Fig. 7(b-d), the rising trend in the second half of the curves for PCC, SSIM, and PSNR is very slow. The results of super-resolution imaging with the speckle size of 256×256 were compressed to 64×64 with PCC, SSIM, and PSNR improved only 1.7%, 1.9%, and 0.7 dB, respectively, compared to the speckle size of 128×128 were compressed to 64×64. In addition, upsampling in Tensorflow interpolates the data in the directions of height and width, resulting in images that are simply scaled up. However, the redundancy of the speckle is not exploited, and the final reconstruction needs to be improved. PixleShuffle maps the feature maps of speckle patterns into high-resolution feature maps by convolution and multi-channel reconfiguration. The utilization of the physical information perceived by the speckle is improved. It helps U-net to use the redundancy of speckle information fully. The hidden physical information is perceived, and the HR images are better reconstructed than that with upsampling, as shown in Fig.8. To visualize the results, the error row shows the prediction of U-net, which is overlaid with true positives (white), false positives (green) and false negatives (purple). The high-resolution images reconstructed using PixleShuffle have only a few reconstruction errors, and the target details are well-preserved. However, when using upsampling, the reconstruction of HR images relies on data-driven only. The details of the target are missing more than that reconstruction with PixleShuffle, relatively.The quantitative results for both are shown in Table 5. The PCC, SSIM, and PSNR of U-net with PixleShuffle are 2.0%, 1.4%, and 0.4 dB higher than those with upsampling, respectively. Finally, the issue of the number of pixels in the speckle pattern needs to focus. We can find that there is a limitation on the physical up-sampling. As demonstrated in the experiment, we performed super-resolution imaging with a maximum magnification of 8. The reason is that the size of the speckle pattern captured by CMOS is less than 512×512. In simple terms, the limited size of the speckle pattern limited the paper's approach.

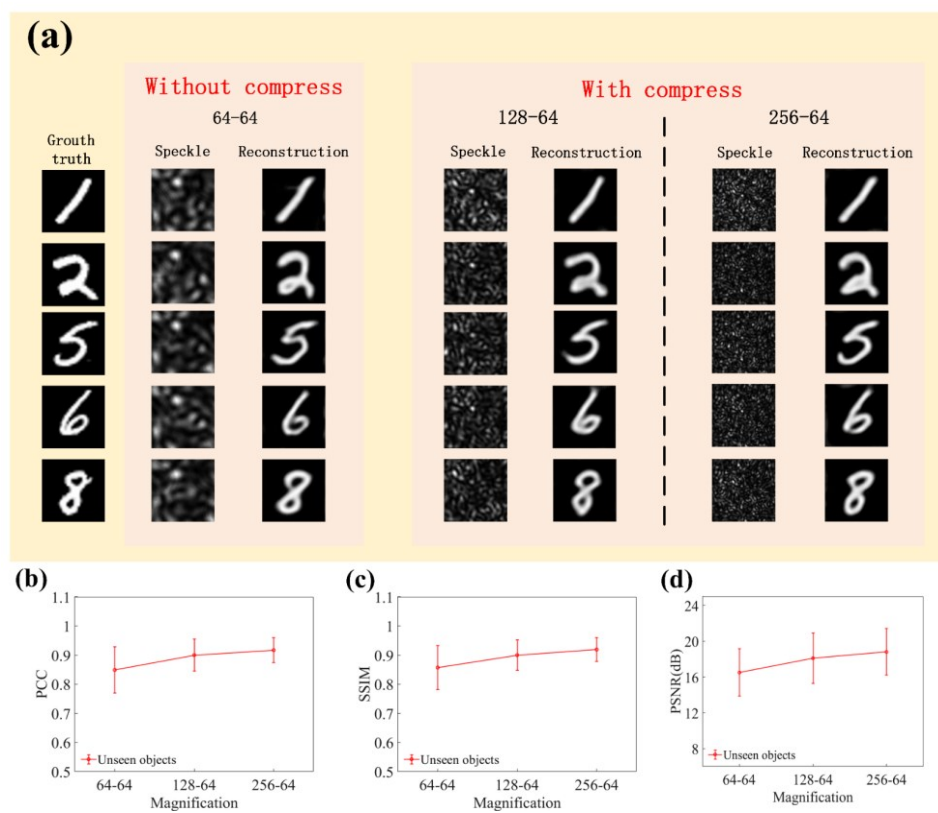

Figure 7 Results of super-resolution imaging with a combination of compressed sensing and speckle-driven. (b), (c) and (d) is the curve chart of averaged PCC, SSIM, and PSNR with error bars, respectively. 64-64: the speckle of size 64×64 is compressed to 64×64, i.e., without compression; 64-64: the speckle of size 128×128 is compressed to 64×64; 256-64: the speckle of size 256×256 is compressed to 64×64.

Table 4. Quantitative evaluation of the physical up-sampling with compressed sensing

|  | PCC | SSIM | PSNR(dB) |
|---|---|---|---|
| 256-64 | 91.7% | 91.9% | 18.8 |
| 128-64 | 90.0% | 90.0% | 18.1 |
| 64-64 | 84.9% | 85.7% | 16.5 |

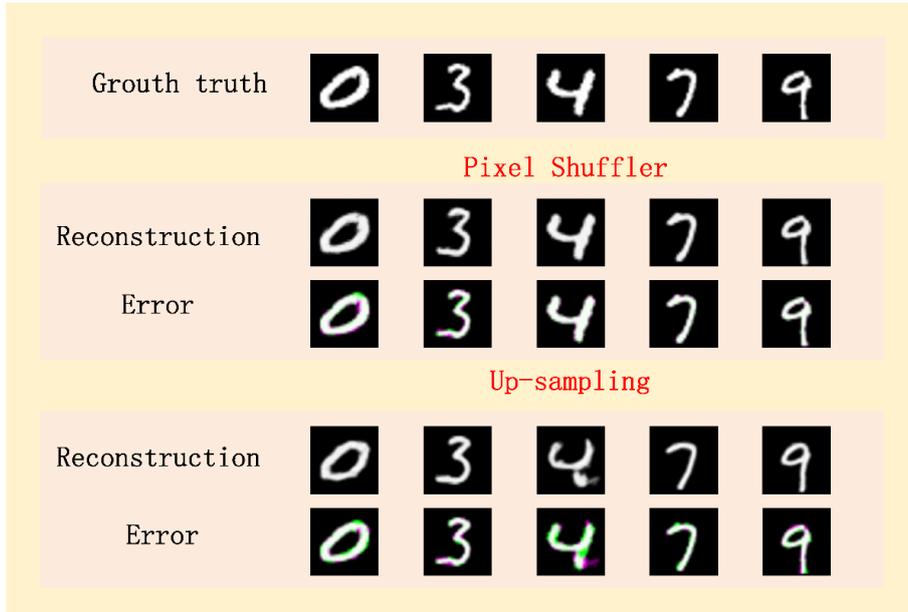

Figure 8 Reconstruction results of unseen objects using upsampling and Pixelshuffle, respectively

Table 5. Quantitative evaluation of reconstruction of unseen objects with upsampling and PixelShuffle

|  | PCC | SSIM | PSNR(dB) |
|---|---|---|---|
| Pixel Shuffle | 97.4% | 94.9% | 21.2 |
| Up-sampling | 95.4% | 93.5% | 20.8 |

# 5. CONCLUSION

In this paper, we explore a method of the physical upsampling of speckle-driven to help super-resolution imaging through MMF. It takes advantage of the information redundancy of the speckle to achieve physical up-sampling. Experimental demonstrations have been provided for the setup of the proposed method, data-driven and the combination of compressed sensing and data-driven, as well as their experimental results. The experimental results confirm the advantages of physical upsampling of speckle-driven to avoid the disadvantages of missing information due to data drive. Significantly, it breaks the limitation of super-resolution imaging through MMF. The quality of HR images does not degrade with increasing magnification. At the same time, it is confirmed that this paper's method, combined with simple compression, can further exploit the advantages of speckle redundancy. Therefore, our approach could be the key to developing minimally invasive surgery. Despite these advantages, if the camera cannot capture a speckle pattern with sufficient size, the advantages of the method in the paper are somewhat controversial. Next, the balance between the enrichment of redundancy and the loss of information by a high compression of speckle needs to be considered. Finally, the reconstruction is worse at low magnification than some methods, such as compression perception. We will follow up with related studies. Some laws are considered for exploration in the subsequent work: 1. explore the effect of the transmission model of MMF on the formation of speckles; 2. study the relation between the redundancy of the speckle and the degree of compression of the speckle pattern. Our experimental results push the boundaries of super-resolution imaging through a multimode fiber.

**Funding.** This work was supported by Fundamental Research Funds for the Central Universities (Grant NO. JSGP202202, 30919011401,30920010001), Leading Technology of Jiangsu Basic Research Plan (BK20192003), Key Research & Development programs in Jiangsu China (Grant no. BE2018126), the Postgraduate Research & Practice Innovation Program of Jiangsu Province (KYCX22_0411, KYCX23_0436), the Open Foundation of Key Lab of Optic-Electronic and Communication of Jiangxi Province (NO.20212OEC002).

**Acknowledgments.** We thank Yicheng Zhou and Zhengtao Song for useful discussions.

**Disclosures.** The authors declare that they have no known competing financial interests or personal relationships that could have appeared to influence the work reported in this paper.

**Data availability.** Data underlying the results presented in this paper are not publicly available but may be obtained from the authors upon reasonable request.

References

1. Y. Liu, M. Yang, J. Zhang, X. Zhi, C. Li, C. Zhang, F. Pan, K. Wang, Y. Yang, J. Martinez de la Fuentea, and D. Cui, "Human Induced Pluripotent Stem Cells for Tumor Targeted Delivery of Gold Nanorods and Enhanced Photothermal Therapy," ACS Nano **10**, 2375-2385 (2016).
2. R. Comesana, F. Quintero, F. Lusquinos, M. J. Pascual, M. Boutinguiza, A. Duran, and J. Pou, "Laser cladding of bioactive glass coatings," Acta Biomater **6**, 953-961 (2010).
3. L. Wang, S. Xie, Z. Wang, F. Liu, Y. Yang, C. Tang, X. Wu, P. Liu, Y. Li, H. Saiyin, S. Zheng, X. Sun, F. Xu, H. Yu, and H. Peng, "Functionalized helical fibre bundles of carbon nanotubes as electrochemical sensors for long-term in vivo monitoring of multiple disease biomarkers," Nat Biomed Eng **4**, 159-171 (2020).
4. A. G. Mignani, and F. Baldini, "Biomedical sensors using optical fibres," Reports on Progress in Physics **59**, 1-28 (1995).
5. M. Rezapour Sarabi, N. Jiang, E. Ozturk, A. K. Yetisen, and S. Tasoglu, "Biomedical optical fibers," Lab Chip **21**, 627-640 (2021).
6. G. Keiser, F. Xiong, Y. Cui, and P. P. Shum, "Review of diverse optical fibers used in biomedical research and clinical practice," J Biomed Opt **19**, 080902 (2014).
7. Y. Chen, A. D. Aguirre, P. L. Hsiung, S. Desai, P. R. Herz, M. Pedrosa, Q. Huang, M. Figueiredo, S. W. Huang, A. Koski, J. M. Schmitt, J. G. Fujimoto, and H. Mashimo, "Ultrahigh resolution optical coherence tomography of Barrett's esophagus: preliminary descriptive clinical study correlating images with histology," Endoscopy **39**, 599-605 (2007).
8. A. Ferlitsch, A. Puespoek, and C. Gasche, "Endoscopic imaging of the vermiform appendix (with video)," Gastrointest Endosc **80**, 1156-1160 (2014).
9. M. Hughes, T. P. Chang, and G. Z. Yang, "Fiber bundle endocytoscopy," Biomed Opt Express **4**, 2781-2794 (2013).
10. S. Sivankutty, V. Tsvirkun, G. Bouwmans, D. Kogan, D. Oron, E. R. Andresen, and H. Rigneault, "Extended field-of-view in a lensless endoscope using an aperiodic multicore fiber," Opt Lett **41**, 3531-3534 (2016).
11. A. Porat, E. R. Andresen, H. Rigneault, D. Oron, S. Gigan, and O. Katz, "Widefield lensless imaging through a fiber bundle via speckle correlations," Opt Express **24**, 16835-16855 (2016).
12. I. T. Leite, S. Turtaev, D. E. Boonzajer Flaes, and T. Čižmár, "Observing distant objects with a multimode fiber-based holographic endoscope," APL Photonics **6** (2021).
13. S. Singh, S. Labouesse, and R. Piestun, "Multiview Scattering Scanning Imaging Confocal Microscopy Through a Multimode Fiber," IEEE Transactions on Computational Imaging **9**, 159-171 (2023).
14. E. R. Andresen, S. Sivankutty, V. Tsvirkun, G. Bouwmans, and H. Rigneault, "Ultrathin endoscopes based on multicore fibers and adaptive optics: a status review and perspectives," J Biomed Opt **21**, 121506 (2016).
15. A. M. Caravaca-Aguirre, and R. Piestun, "Single multimode fiber endoscope," Opt Express **25**, 1656-1665 (2017).
16. C. Zhang, Z. Yao, Z. Qin, G. Gu, Q. Chen, Z. Xie, G. Liu, and X. Sui, "Optical refocusing through perturbed multimode fiber using Cake-Cutting Hadamard encoding algorithm to improve robustness," Optics and Lasers in Engineering **164** (2023).
17. I. M. Vellekoop, "Feedback-based wavefront shaping," Opt Express **23**, 12189-12206 (2015).
18. T. Cizmar, and K. Dholakia, "Shaping the light transmission through a

multimode optical fibre: complex
transformation analysis and applications
in biophotonics," OPTICS EXPRESS **19**, 18871-18884 (2011).
19. S. Resisi, Y. Viernik, S. M. Popoff, and Y. Bromberg, "Wavefront shaping in multimode fibers by transmission matrix engineering," APL Photonics **5** (2020).
20. S. Turtaev, I. T. Leite, T. Altwegg-Boussac, J. M. P. Pakan, N. L. Rochefort, and T. Čižmár, "High-fidelity multimode fibre-based endoscopy for deep brain in vivo imaging," Light: Science & Applications **7** (2018).
21. M. Plöschner, T. Tyc, and T. Čižmár, "Seeing through chaos in multimode fibres," Nature Photonics **9**, 529-535 (2015).
22. S. Popoff, G. Lerosey, M. Fink, A. C. Boccara, and S. Gigan, "Image transmission through an opaque material," Nat Commun **1**, 81 (2010).
23. I. N. Papadopoulos, S. Farahi, C. Moser, and D. Psaltis, "High-resolution, lensless endoscope based on digital scanning through a multimode optical fiber," Biomedical Optics Express **4**, 260-270 (2013).
24. I. N. Papadopoulos, S. Farahi, C. Moser, and D. Psaltis, "Focusing and scanning light through a multimode optical fiber using digital phase conjugation," Optics Express **20**, 10583-10590 (2012).
25. L. V. Amitonova, and J. F. de Boer, "Compressive imaging through a multimode fiber," Opt Lett **43**, 5427-5430 (2018).
26. L. G. Wright, D. N. Christodoulides, and F. W. Wise, "Controllable spatiotemporal nonlinear effects in multimode fibres," Nature Photonics **9**, 306-310 (2015).
27. W. Li, K. Abrashitova, G. Osnabrugge, and L. V. Amitonova, "Generative Adversarial Network for Superresolution Imaging through a Fiber," Physical Review Applied **18** (2022).
28. B. Rahmani, I. Oguz, U. Tegin, J.-l. Hsieh, D. Psaltis, and C. Moser, "Learning to image and compute with multimode optical fibers," Nanophotonics **11**, 1071-1082 (2022).
29. P. Fan, T. Zhao, and L. Su, "Deep learning the high variability and randomness inside multimode fibers," Opt Express **27**, 20241-20258 (2019).
30. N. Borhani, E. Kakkava, C. Moser, and D. Psaltis, "Learning to see through multimode fibers," Optica **5** (2018).
31. Z. Liu, L. Wang, Y. Meng, T. He, S. He, Y. Yang, L. Wang, J. Tian, D. Li, P. Yan, M. Gong, Q. Liu, and Q. Xiao, "All-fiber high-speed image detection enabled by deep learning," Nat Commun **13**, 1433 (2022).
32. H. Chen, Z. He, Z. Zhang, Y. Geng, and W. Yu, "Binary amplitude-only image reconstruction through a MMF based on an AE-SNN combined deep learning model," Opt Express **28**, 30048-30062 (2020).
33. B. Rahmani, D. Loterie, G. Konstantinou, D. Psaltis, and C. Moser, "Multimode optical fiber transmission with a deep learning network," Light Sci Appl **7**, 69 (2018).
34. Y. Li, Y. Xue, and L. Tian, "Deep speckle correlation: a deep learning approach toward scalable imaging through scattering media," Optica **5** (2018).
35. B. Song, C. Jin, J. Wu, W. Lin, B. Liu, W. Huang, and S. Chen, "Deep learning image transmission through a multimode fiber based on a small training dataset," Opt Express **30**, 5657-5672 (2022).
36. S. Li, C. Saunders, D. J. Lum, J. Murray-Bruce, V. K. Goyal, T. Cizmar, and D. B. Phillips, "Compressively sampling the optical transmission matrix of a multimode fibre," Light Sci Appl **10**, 88 (2021).

37. S. Zhu, E. Guo, J. Gu, L. Bai, and J. Han, "Imaging through unknown scattering media based on physics-informed learning," Photonics Research **9** (2021).
38. U. Kurum, P. R. Wiecha, R. French, and O. L. Muskens, "Deep learning enabled real time speckle recognition and hyperspectral imaging using a multimode fiber array," Opt Express **27**, 20965-20979 (2019).
39. E. Guo, C. Zhou, S. Zhu, L. Bai, and J. Han, "Dynamic imaging through random perturbed fibers via physics-informed learning," Optics & Laser Technology **158** (2023).
40. Q. Cheng, L. Bai, J. Han, and E. Guo, "Super-resolution imaging through the diffuser in the near-infrared via physically-based learning," Optics and Lasers in Engineering **159** (2022).
41. X. Zhang, S. Cheng, J. Gao, Y. Gan, C. Song, D. Zhang, S. Zhuang, S. Han, P. Lai, and H. Liu, "Physical origin and boundary of scalable imaging through scattering media: a deep learning-based exploration," Photonics Research **11** (2023).
42. X. Zhang, J. Gao, Y. Gan, C. Song, D. Zhang, S. Zhuang, S. Han, P. Lai, and H. Liu, "Different channels to transmit information in scattering media," PhotoniX **4** (2023).
43. G. Barbastathis, A. Ozcan, and G. Situ, "On the use of deep learning for computational imaging," Optica **6** (2019).
44. S. M. Popoff, G. Lerosey, M. Fink, A. C. Boccara, and S. Gigan, "Controlling light through optical disordered media: transmission matrix approach," New Journal of Physics **13** (2011).
45. "https://github.com/pankajvshrma/PixelShuffle-Tensorflow/blob/master/pixelshuffle.py."
46. W. Z. Shi, J. Caballero, F. Huszar, J. Totz, A. P. Aitken, R. Bishop, D. Rueckert, and Z. H. Wang, "Real-Time Single Image and Video Super-Resolution Using an Efficient Sub-Pixel Convolutional Neural Network," 2016 Ieee Conference on Computer Vision and Pattern Recognition (Cvpr), 1874-1883 (2016).
47. S.-J. Kang, "SSIM Preservation-Based Backlight Dimming," Journal of Display Technology **10**, 247-250 (2014).
48. I. Bakurov, M. Buzzelli, R. Schettini, M. Castelli, and L. Vanneschi, "Structural similarity index (SSIM) revisited: A data-driven approach," Expert Systems with Applications **189** (2022).
49. Z. Wang, A. C. Bovik, H. R. Sheikh, and E. P. Simoncelli, "Image quality assessment: from error visibility to structural similarity," IEEE Trans Image Process **13**, 600-612 (2004).